\documentclass[preprint,a4paper,twocolumn,bibnotes,10pt,aps]{revtex4}%
\usepackage{amsfonts}
\usepackage{amsmath}
\usepackage{amssymb}
\usepackage{graphicx}%
\begin{document}
\title{Anomalous fluctuation regimes at the FFLO transition }
\date{\today}
\author{Fran\c{c}ois Konschelle}
\author{J\'{e}r\^{o}me Cayssol }
\author{Alexandre I. Buzdin}
\altaffiliation{Also at Institut Universitaire de France}

\affiliation{Universit\'{e}s de Bordeaux ; CPMOH ; CNRS, 33405 Talence, France}


















































































\begin{abstract}
Recently some experimental evidences have been obtained in favour of the
existence of the inhomogeneous Fulde-Ferrell-Larkin-Ovchinnikov (FFLO)
superconducting state in heavy-fermion superconductor CeCoIn$_{5}$ and organic
superconductor $\lambda$-(BETS)$_{2}$FeCl$_{4}$. However the unambiguous
identification of FFLO state remains very difficult. We present the
theoretical studies of the Gaussian fluctuations near the tricritical point
(where the FFLO modulation appears) and demonstrate that the behavior of the
fluctuational specific heat, paraconductivity and diamagnetism is
qualitatively different from the usual superconducting transition. Special
values of the critical exponent and the crossovers between different
fluctuational regimes may provide a unique test for the FFLO state appearance.

\end{abstract}
\startpage{1}
\maketitle

\bigskip

\section{Introduction}

The long-standing hunt for the inhomogenous superconducting state predicted by
Fulde and Ferrell \cite{fulde_ferrell(1964)} and by Larkin and Ovchinnikov
\cite{larkin_ovchinnikov(1964)} (the so-called FFLO state) has recently known
a revival in the study of fermionic cold atoms
\cite{zwierlein(science).2006,partridge.2006}, quantum chromodynamics color
superconductivity \cite{casalbuoni.nardulli.2004} and heavy fermion
superconductor CeCoIn$_{5}$
\cite{radovan_murphy.2003,bianchi_movshovich.2002,miclea.nicklas.2006}. The
FFLO phase consists in a condensate of finite momentum Cooper pairs in
contrast to the zero momenta pairs of the usual BCS state. Hence the FFLO
superconducting order parameter acquires a spatial variation. Such a state may
be induced when a chemical potential difference is applied between two species
of fermions with mutual pairing attraction.

In superconductors the different chemical potentials are obtained by a Zeeman
splitting $h=\mu_{B}H$ between spin up and down states, $\mu_{B}$ and $H$
being respectively the Bohr magneton and an external magnetic field (in
magnetic superconductors $h$ is the internal exchange field). A characteristic
feature of the field-temperature phase diagram is the existence of a
tricritical point (TCP) which is the meeting point of three transition lines
separating the normal metal, the uniform superconductor and the FFLO state.
For clean $s$-wave superconductors the TCP is located at $T^{\ast}=0.56T_{c0}$
and $h^{\ast}=1.07k_{B}T_{c0}$, $T_{c0}$ being the zero field critical
temperature. Unfortunately, the FFLO state is only energetically favorable in
a small part of the phase diagram, located at low temperatures $T<T^{\ast}$
and high fields, see Fig.\ref{FIG_phase_diagram}. Up to now, there is no
compelling experimental evidence for this mysterious FFLO phase although some
heavy fermion superconductors like CeCoIn$_{5}$
\cite{radovan_murphy.2003,bianchi_movshovich.2002,miclea.nicklas.2006},
organic superconductors like $\lambda$-(BETS)$_{2}$FeCl$_{4}$
\cite{balicas.2001,uji_graf_brooks.2006}, and rare-earth magnetic
superconductors like ErRh$_{4}$B$_{4}$ \cite{bulaevskii_buzdin.1985} are
promising candidates.

In the neighborhood of the TCP, the FFLO transition may be described within a
modified Ginzburg-Landau approach (MGL). The MGL free energy functional, which
follows from BCS theory with Zeeman pair breaking, was first obtained for
$s$-wave superconductors \cite{buzdin_kachkachi(1997)} and then generalized to
$d$-wave pairing with/without impurities \cite{yang_agterberg.2001}. The
salient features of the MGL functional are the presence of higher order
derivatives of the order parameter $\Psi$ than the usual $g\left\vert
\mathbf{\nabla}\Psi\right\vert ^{2}$ term, and the fact that the coefficient
$g$ of the latter term changes sign at the TCP being positive for $h/2\pi
k_{B}T<0.3$ and negative otherwise. When the stiffness $g$ becomes negative,
the spatially varying configurations of the order parameter are favored
leading to the FFLO transition. As a result of the anisotropy contained in
higher order derivatives, the degeneracy over the FFLO modulation is removed
even in the cubic lattice. Another source of anisotropy is related to
momentum-dependent pairing. In the case of a $d$-wave Pauli limited
superconductor both lattice and pairing anisotropies are of primary importance
for the FFLO \cite{yang_sondhi.1998,vorontsov_sauls_graf.2005,shimahara.2002}.

Beside these mean-field studies, the fluctuations have rarely been considered
in the context of the FFLO transition, although they may be accounted for in a
quite simple and accurate manner within the framework of the MGL. Using the
standard Ginzburg-Landau functional, the effect of fluctuations on the
properties of the normal state were extensively studied for the homogeneous
BCS superconductors \cite{skocpol_tinkham.1975,b.tinkham} and high-$T_{c}$
superconductors \cite{b.larkin_varlamov}. Though relatively small,
thermodynamical and transport properties were actually measured in the
Gaussian regime.

Up to now, a detailed investigation of the fluctuations near the FFLO
transition was missing. A first attempt to fill this flaw was realized
recently by Marenko and Samokhin who addressed the issue of the quantum
critical fluctuations at the zero temperature FFLO transition
\cite{samokhin.marenko.2006}. The aim of this paper is to perform the
complementary investigation of the fluctuations in the neighborhood of the
tricritical point.

In this Letter, we study the effect of the fluctuations close to the FFLO
transition. We focus on the specific heat $C$ and on the excess conductivity
$\sigma$, the so-called paraconductivity, induced by the superconducting
fluctuations within the normal state. We have identified three regimes in the
vicinity of the tricritical point $\left(  T^{\ast},h^{\ast}\right)  $ which
differ by the nature of the soft fluctuation modes.

Within the first regime, depicted as region I in Fig.\ref{FIG_phase_diagram},
the soft modes have the usual quadratic dispersion and are located around the
origin of momentum space.\ This familiar situation yields the well-established
power law behaviors $\left(  C,\sigma\right)  \sim\left(  T-T_{c}\right)
^{(d-4)/2}$ reviewed in
\cite{skocpol_tinkham.1975,b.tinkham,b.larkin_varlamov}.\begin{figure}[ptb]
\begin{center}
\includegraphics[angle=0,width=3in]{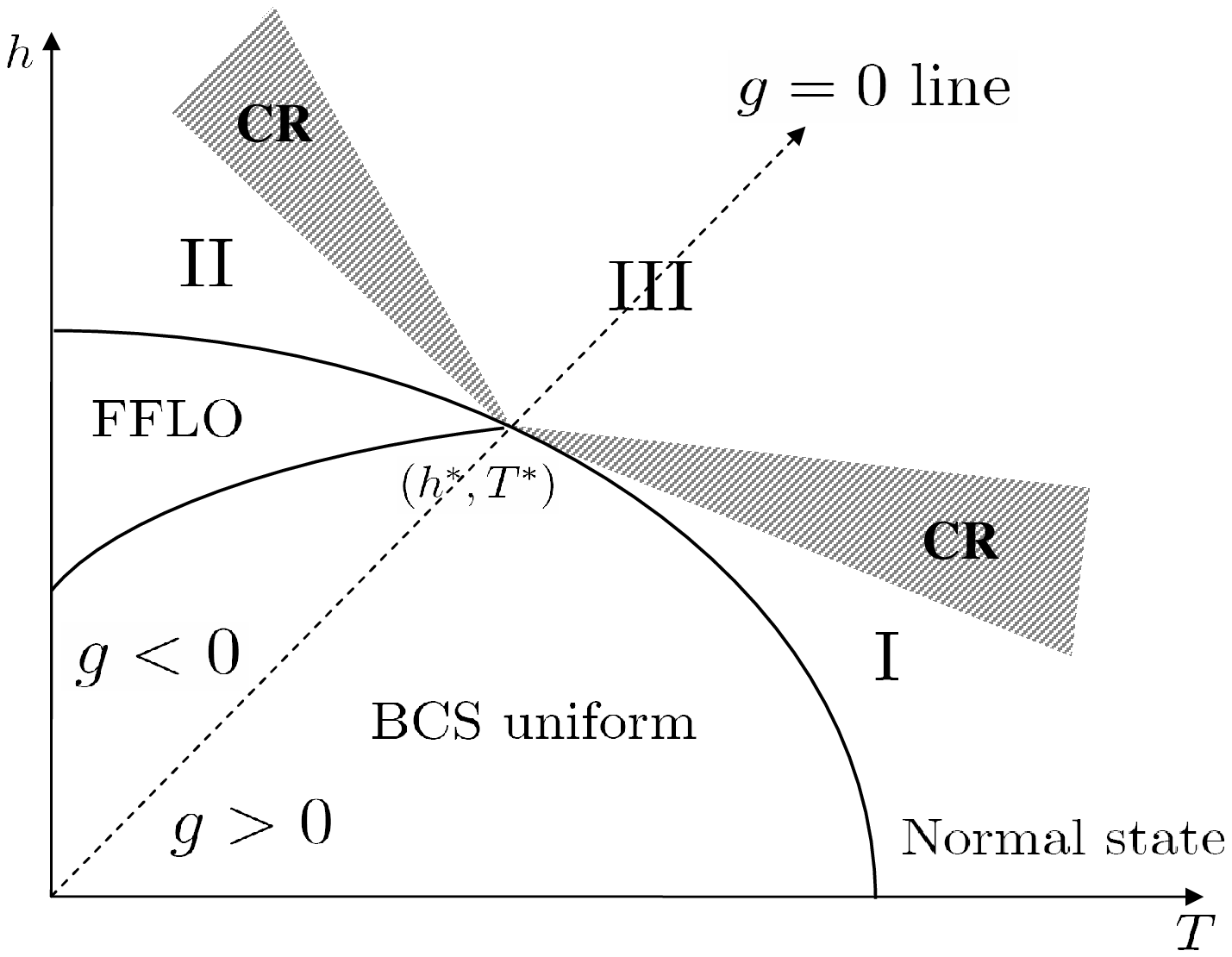}
\end{center}
\caption{Schematic $(h,T)$ phase diagram around the tricritical point
$(h^{\ast},T^{\ast})$. The solid lines represent the transitions between the
normal metal, the spatially uniform BCS superconductor, and the non uniform
FFLO state. In region I where $g>0$ (resp. region II where $g<0$ ), the normal
state behavior is dominated by uniform (resp. non uniform) superconducting
modes which propagate quadratically. In region III where $g\rightarrow0$, the
preponderant fluctuations propagate as $\varepsilon_{\mathbf{k}}\sim k^{4}$.
Region III is defined by $\left\vert T-T_{c}\right\vert \gg g^{2}/a\gamma.$
There are also crossover regimes (CR) separating the region III from I and II.
}%
\end{figure}

In contrast, the second regime represented as region II in
Fig.\ref{FIG_phase_diagram}, is characterized by anomalous power laws in
isotropic or weakly anisotropic materials. Here anomalous simply means
distinct from $\left(  C,\sigma\right)  \sim\left(  T-T_{c}\right)
^{(d-4)/2}$. This discrepancy is related to the fact that the minima of the
dispersion are attained on a finite dimensional surface, a sphere or a ring,
see Fig.\ref{FIG_crossovers}.$\left(  c\right)  $. Hence the available phase
space volume of the low energy fluctuations is larger than in the case of the
standard BCS transition towards a uniform ground state, as it was first
pointed out by Brazovski \cite{brazovski}. On the other hand, among the host
of promising candidates for the FFLO state, crystal or gap symmetry effects
are important. In such case the sphere or ring containing the low energy modes
is partially removed and the low energy fluctuations are located within small
islands centered at a few number of isolated points in momentum space, as it
is shown in Fig.\ref{FIG_crossovers}.$\left(  a\right)  $. Thus for large
anisotropy, the anomaly disappears and one recovers the usual power laws
$\left(  C,\sigma\right)  \sim\left(  T-T_{c}\right)  ^{(d-4)/2}$. In summary
the exponents in regime II depend drastically on the anisotropy of the system.

The most interesting regime, region III of Fig.\ref{FIG_phase_diagram} is at
the immediate vicinity of the TCP where the FFLO and the uniform
superconducting phases compete. The power laws of the specific heat
$C\sim\left(  T-T_{c}\right)  ^{(d-8)/4}$ and paraconductivity $\sigma
\sim\left(  T-T_{c}\right)  ^{(d-6)/4}$ are anomalous and quite universal
since they pertain in both isotropic and anisotropic models. These predictions
rely on the fact that the dispersion of the soft fluctuation modes becomes
quartic within this regime.

Indeed, these three regimes and the related crossovers between them may serve
as a powerfull tool to identify FFLO state.

\section{General formalism}

We investigate the fluctuation properties at the transition between normal and
FFLO states using a quite general approach based on the MGL functional
\cite{buzdin_kachkachi(1997),yang_agterberg.2001}. Note that we assume a
second order phase transition to the FFLO state. In the framework of isotropic
model this transition is of second order for one and two dimensional systems
while it is of first order in three dimensional case
\cite{buzdin_kachkachi(1997)}. However in the real superconductors the
crystalline anisotropy (even a cubic one) is of primary importance and can
modify this conclusion. Nevertheless, we may expect that in the case of weakly
first order FFLO transition, the fluctuation regimes studied here would be
still observable. Such description is adequate near the tricritical point
where the most striking behavior is expected and where the wave-vector of the
FFLO modulation is small. Looking for Gaussian fluctuations in the normal
state it is sufficient to consider MGL functional only in the quadratic
$\left\vert \Psi_{\mathbf{k}}\right\vert ^{2}$ approximation
\begin{equation}
H\left[  \Psi\right]  =\sum_{\mathbf{k}}\text{ }\underset{i,j=1}{%
{\displaystyle\sum^{d}}
}\left(  \alpha+gk_{i}^{2}+\gamma_{ij}k_{i}^{2}k_{j}^{2}\right)  \left\vert
\Psi_{\mathbf{k}}\right\vert ^{2} \label{EQ_our_model}%
\end{equation}
where $\Psi_{\mathbf{k}}$ are the Fourier components of the complex order
parameter and $d$ the dimensionality of the sample. The coefficient
$\alpha=a(T-\widetilde{T}_{c})$, the coefficients $g$ and $\gamma_{ij}$ depend
on the magnetic field and temperature, $\widetilde{T}_{c}\left(  h\right)  $
being the critical temperature for the second order transition between normal
and uniform superconducting states. Although qualitative conclusions are quite
general, explicit calculations are performed within the cubic anisotropy model
defined by $\gamma_{ij}=\gamma\delta_{ij}+\gamma\eta\left(  1-\delta
_{ij}\right)  $, where $\gamma>0$. The case $\eta=1$ corresponds to the
isotropic model since then the general fluctuation spectrum $\varepsilon
_{\mathbf{k}}=$ $\alpha+gk^{2}+\Sigma_{i,j}\gamma_{ij}k_{i}^{2}k_{j}^{2}$
depends only on $k^{2}=\Sigma_{i}k_{i}^{2}$ and on $k^{4}$. Hence $\left\vert
\eta-1\right\vert $ is a measure of the deviation from the ideally isotropic
material with a spherical Fermi surface.

Integrating out the Gaussian modes $\Psi_{\mathbf{k}}$, one obtains the
fluctuation free energy density
\begin{equation}
F=k_{B}T\frac{1}{L^{d}}\sum_{\mathbf{k}}\ln\frac{\varepsilon_{\mathbf{k}}}{\pi
k_{B}T} \label{Free}%
\end{equation}
where $L$ is the size of the system. The specific heat may be readily deduced
from the free energy (\ref{Free}) via
\begin{equation}
C=-T_{c}\frac{\partial^{2}F}{\partial T^{2}}=a^{2}k_{B}T_{c}^{2}\frac{1}%
{L^{d}}\sum_{\mathbf{k}}\text{ }\frac{1\text{ }}{\varepsilon_{\mathbf{k}}^{2}%
}\text{ \ .} \label{EQ_heat_capa}%
\end{equation}
In evaluating the paraconductivity within the Kubo formalism, one should use
the time-dependent Ginzburg-Landau equation to obtain the current-current
correlator at different times\cite{b.tinkham}. Moreover attention should be
paid to the fact that, owing to the presence of high order derivatives
$\gamma_{ij}k_{i}^{2}k_{j}^{2}$ in Eq.(\ref{Free}), the expression for the
current operator differs from the usual one. However the general
expression\cite{b.larkin_varlamov} relating the paraconductivity to the
fluctuation spectrum,%
\begin{equation}
\sigma_{xx}=\sigma_{yy}=\frac{\pi e^{2}ak_{B}T_{c}}{4\hslash}\frac{1}{L^{d}%
}\sum_{\mathbf{k}}\frac{v_{\mathbf{k}x}^{2}}{\varepsilon_{\mathbf{k}}^{3}},
\label{EQ_sigmagene}%
\end{equation}
still holds for the FFLO case although expressions for the velocities
$v_{\mathbf{k}x}=\partial\varepsilon_{\mathbf{k}}/\partial k_{x}$ and for the
current are very different from the usual quantum mechanical formula.

Owing to the denominators in Eqs.(\ref{EQ_heat_capa},\ref{EQ_sigmagene}) the
main contributions to the specific heat and conductivity originate from the
modes with the lowest energies $\varepsilon_{\mathbf{k}}$. The above
expressions lead to singular power laws which diverge at the transition with
characteristic exponents. The exponents depend on the nature of the low energy
modes: their location in momentum space and their dispersion. In the usual
case of a uniform superconducting state, namely for $g>0$ and $\gamma_{ij}=0,$
the low energy modes are concentrated around the center of the Brillouin zone
and propagate with a quadratic dispersion. The resulting power-law
divergencies of the thermodynamical and transport properties are well-known
\cite{skocpol_tinkham.1975,b.tinkham,b.larkin_varlamov}. In the vicinity of
the tricritical point, the situation is much more interesting since three
different kinds of spectra $\varepsilon_{\mathbf{k}}$, namely
(\ref{EQ_spectreiso}), (\ref{EQ_spectreaniso}) or (\ref{EQ_spectrum_g_0}),
lead to different power laws.\begin{figure}[ptb]
\begin{center}
\includegraphics[angle=0,width=3in]{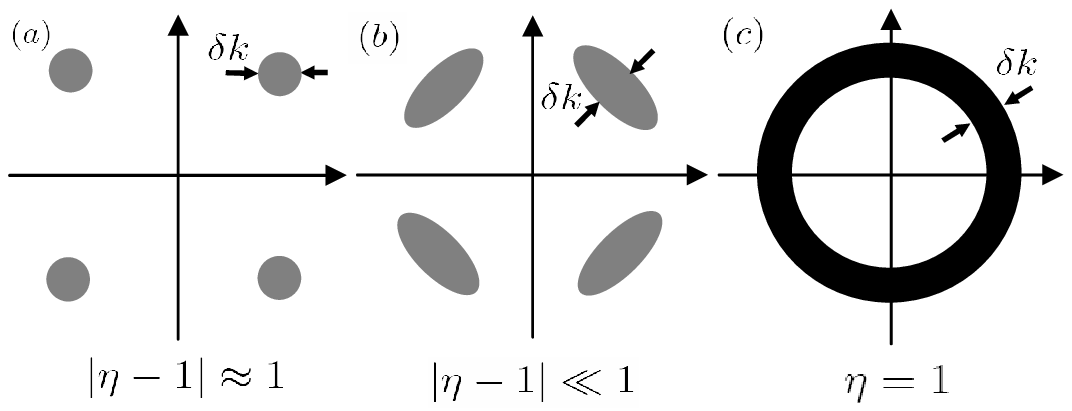}
\end{center}
\caption{Crossover between isotropic $\left(  c\right)  $ and anisotropic
$\left(  a\right)  $ and $\left(  b\right)  $ models in the regime $\left\vert
T-T_{c}\right\vert \ll g^{2}/a\gamma$. Shaded areas represent the low energy
modes leading to the various power laws.}%
\end{figure}

\section{Isotropic model and large stiffness}

We first consider the isotropic case $\eta=1$. At low fields and high
temperatures $h/2\pi k_{B}T<0.3$, the positive stiffness $g$ leads to the
standard spatially uniform superconducting state (referred as BCS state on
Fig.\ref{FIG_phase_diagram}). Conversely when $h/2\pi k_{B}T>0.3$, the
stiffness becomes negative and the fluctuation spectrum $\varepsilon
_{\mathbf{k}}^{iso}=\alpha+gk^{2}+\gamma k^{4}$ may be rewritten as
\begin{equation}
\varepsilon_{\mathbf{k}}^{iso}=\tau+\gamma\left(  k^{2}-q_{0}^{2}\right)
^{2}\text{ \ } \label{EQ_spectrum}%
\end{equation}
where $q_{0}^{2}=-g/2\gamma$ and $\tau=\alpha-g^{2}/4\gamma$. Below
$T_{c}=\widetilde{T}_{c}+g^{2}/4\gamma a\geq\widetilde{T}_{c}$ the normal
state is instable towards the formation of the FFLO state with modulation wave
vector $\boldsymbol{q}_{0}$. Owing to the isotropy of the model, all
directions of $\boldsymbol{q}_{0}$ are degenerate. Above $T_{c}$ the minima of
$\varepsilon_{\mathbf{k}}^{iso}$ are attained on a finite dimensional surface:
a sphere for $d=3$ or a ring for $d=2,$ whose radius is exactly $q_{0}.$ The
one dimensional case is special in the sense that the minima consist in two
points with coordinates $\pm q_{0}$.

At the vicinity of the sphere/ring defined by $k=q_{0}$ the approximation%
\begin{equation}
\varepsilon_{\mathbf{k}}^{iso}=a\left(  T-T_{c}\right)  +4\gamma q_{0}%
^{2}\left(  k-q_{0}\right)  ^{2}\text{ \ } \label{EQ_spectreiso}%
\end{equation}
may be used to evaluate Eqs.(\ref{EQ_heat_capa},\ref{EQ_sigmagene}) leading to
the following singular behaviors for the specific heat
\begin{equation}
~C_{iso}=A_{d}k_{B}T_{c}^{2}\left(  \frac{a}{\gamma}\right)  ^{1/2}\left(
\frac{\left\vert g\right\vert }{2\gamma}\right)  ^{\left(  d-2\right)
/2}\left(  T-T_{c}\right)  ^{-3/2}~~~~, \label{Heatiso}%
\end{equation}
and for the paraconductivity%

\begin{equation}
\sigma_{iso}=\frac{\pi A_{d}e^{2}k_{B}T_{c}}{\hslash d}\left(  \frac{\gamma
}{a}\right)  ^{1/2}\left(  \frac{\left\vert g\right\vert }{2\gamma}\right)
^{d/2}\left(  T-T_{c}\right)  ^{-3/2}, \label{Condiso}%
\end{equation}
with $A_{d}=1/4,1/8,1/8\pi$ for $d=1,2,3$ respectively. Both $C_{iso}$ and
$\sigma_{iso}$ diverge at the FFLO critical temperature $T_{c}$ with the same
$d$-independent exponent in contrast to the usual uniform superconductivity
case. In particular for the quasi-two and three dimensional cases, the
divergencies of the fluctuation contributions near the FFLO transition are
much stronger than the $\left(  T-T_{c}\right)  ^{(d-4)/2}$ power law
characterizing the usual superconducting transition. In this sense, the
anomalous behavior of the Gaussian fluctuations is a signature of the FFLO
state in such ideal isotropic systems.

\section{Anisotropic models and large stiffness}

The previous results $\left(  \ref{Heatiso},\ref{Condiso}\right)  $ rely
deeply on the perfect isotropy of the model. Nevertheless real superconductors
are anisotropic owing to either crystalline effect or $d$-wave pairing. Hence
the minima of the dispersion occur at a few number $N_{\alpha\text{ }}$of
isolated points $\boldsymbol{q}_{0}^{(\alpha)}$ which are located at a finite
distance from the origin. These points belong to high symmetry lines of the
crystal. More specifically in the model $\gamma_{ij}=\gamma\delta_{ij}%
+\gamma\eta\left(  1-\delta_{ij}\right)  $ the wave vectors $\boldsymbol{q}%
_{0}^{(\alpha)}$ are along the diagonals when $\eta<1$ and along the
crystalline axis for $\eta>1$, whereas $\eta=1 $ corresponds to the isotropic
case. The relevant low energy fluctuations form small islands around the
points $\boldsymbol{q}_{0}^{(\alpha)}$, as shown in Fig.\ref{FIG_crossovers}%
.$\left(  a\right)  $. In the case of a very weak anisotropy $\left\vert
\eta-1\right\vert \ll1$, these pockets may collapse within a finite
dimensional surface and one recovers the isotropic model discussed above, see
Fig.\ref{FIG_crossovers}.$\left(  b,c\right)  $.

Here we consider the opposite case of significant anisotropies $\left\vert
\eta-1\right\vert \sim1$ which is expected in real compounds. Then the favored
modes belong to $N_{\alpha\text{ }}$disconnected islands. Hence one may focus
on each of the $\boldsymbol{q}_{0}^{(\alpha)}$ obtained by solving
$\partial_{i}\varepsilon_{\mathbf{k}}(\boldsymbol{q}_{0}^{(\alpha)})=0$ and
expand the fluctuation spectrum of $\left(  \ref{EQ_our_model}\right)  $
around $\boldsymbol{q}_{0}^{(\alpha)}$ as%
\begin{equation}
\varepsilon_{\mathbf{k}}^{ani}=a\left(  T-T_{c}\right)  +\underset{i,j=1}{%
{\displaystyle\sum^{d}}
}\lambda_{ij}^{(\alpha)}\left(  k_{i}-q_{0i}^{(\alpha)}\right)  \left(
k_{j}-q_{0j}^{(\alpha)}\right)  \label{EQ_spectreaniso}%
\end{equation}
where $\lambda_{ij}^{(\alpha)}=\left(  1/2\right)  (\partial_{i}%
\varepsilon_{\mathbf{k}}\partial_{j}\varepsilon_{\mathbf{k}})(\boldsymbol{q}%
_{0}^{(\alpha)})$ is the effective mass tensor for the fluctuating modes
located around $\boldsymbol{q}_{0}^{(\alpha)}$. Owing to the symmetry of the
crystal, the eigenvalues $\lambda_{i}$ of the tensors $\lambda_{ij}^{(\alpha
)}$ are the same for the $N_{\alpha}$ islands.

Evaluating $\left(  \ref{EQ_heat_capa}\right)  $ and $\left(
\ref{EQ_sigmagene}\right)  $ with the spectrum $\left(  \ref{EQ_spectreaniso}%
\right)  $ and collecting the contributions from the $N_{\alpha\text{ }}%
$minima yield the specific heat
\begin{equation}
C_{ani}\simeq N_{\alpha}k_{B}T_{c}^{2}\frac{a^{d/2}}{\sqrt{\det\lambda_{ij}}%
}\left(  T-T_{c}\right)  ^{\left(  d-4\right)  /2}\text{ } \label{HeatAniso}%
\end{equation}
and the diagonal paraconductivity $\sigma_{ani}(=\sigma_{xx}=\sigma_{yy})$ in
the plane $xy$
\begin{equation}
\sigma_{ani}\simeq\frac{N_{\alpha}}{2}\frac{e^{2}k_{B}T_{c}}{\hslash}\left(
\frac{(\lambda_{1}+\lambda_{2})a^{d/2-1}}{\sqrt{\det\lambda_{ij}}}\right)
\left(  T-T_{c}\right)  ^{\left(  d-4\right)  /2} \label{CondAniso}%
\end{equation}
where $\lambda_{1}$ and $\lambda_{2}$ are the eigenvalues of $\lambda
_{ij}^{(\alpha)}$ whose eigenvectors span the $xy$ plane. For the conductivity
$\sigma_{zz}$ one should replace $\lambda_{1}+\lambda_{2}$ by $2\lambda_{z}$.
In sharp contrast to the isotropic model results (\ref{Heatiso},\ref{Condiso}%
), the temperature dependence of both $C_{ani}$ and $\sigma_{ani}$ becomes
sensitive to the dimension with the same exponents as in the homogeneous case
\cite{skocpol_tinkham.1975,b.larkin_varlamov}. Thus in the anisotropic model,
the exponents themselves no longer provide a valuable hallmark for the FFLO
state. Naturally when $\eta\rightarrow1$ one should recover the isotropic
results (\ref{Heatiso},\ref{Condiso}). The crossover from anisotropic to
isotropic models is provided by the $\eta$ dependence of $\det\lambda_{ij}.$
The latter determinant vanishes in the limit $\eta\rightarrow1$, as an example
like $\det\lambda_{ij}=g^{2}(1-\eta^{2})$ in the quasi-two dimensional case.

\section{Three fluctuation regimes}

Let us check now the validity of the approximations (\ref{EQ_spectreiso}) and
(\ref{EQ_spectreaniso}) used so far in deriving the specific heat and
paraconductivity. We aim to identify the region of the phase diagram where the
previous results (\ref{Heatiso},\ref{Condiso}) and (\ref{HeatAniso}%
,\ref{CondAniso}) apply. The modes which participate to the singular parts of
the specific heat and conductivity are spread within a shell of width $\delta
k$ around the circle $k=q_{0}$ (isotropic case) or in islands of size $\delta
k$ centered at isolated points (anisotropic case). The width $\delta k$, which
is defined by the condition $4q_{0}^{2}\gamma\delta k^{2}\approx\tau
=a(T-T_{c})$,\ must be smaller than $q_{0}$ in order to have well defined
shell or islands (like in Fig.\ref{FIG_crossovers}.$\left(  a\right)  $).
Hence the common range of validity of Eqs.(\ref{Heatiso},\ref{Condiso}) and
(\ref{HeatAniso},\ref{CondAniso}) is given by $\tau\ll\gamma q_{0}^{4}$ or
equivalently $\left\vert T-T_{c}\right\vert \ll g^{2}/a\gamma.$ This
corresponds to the area of the phase diagram which is sufficiently apart from
the tricritical point (large enough $g$) and close to critical line (small
enough $\left\vert T-T_{c}\right\vert $), schematically the region II on
Fig.\ref{FIG_phase_diagram}.

Conversely when $\delta k\gtrsim q_{0}$ the relevant low energy modes merge
towards the center of the Brillouin zone and the approximations
(\ref{EQ_spectreiso}) and (\ref{EQ_spectreaniso}) fail. Hence the specific
heat and paraconductivity must be reevaluated in the regime $\left\vert
T-T_{c}\right\vert \gg g^{2}/a\gamma$. Then the physics is dominated by
extremely soft fluctuation modes having a quartic dispersion instead of the
more usual quadratic one, yielding very special behaviors in the region III of
Fig.\ref{FIG_phase_diagram}. \ 

Finally we investigate the crossover regime between region I and III, see
Fig.\ref{FIG_phase_diagram}. When $g>0$ the specific heat is given by
\begin{equation}
C_{iso}=\frac{a^{2}k_{B}T_{c}^{2}}{\gamma^{d/4}}\alpha^{(d-8)/4}%
{\displaystyle\int\nolimits_{0}^{\infty}}
\frac{dx\Omega_{d}x^{d-1}}{\left(  1+x^{4}+gx^{2}/\sqrt{\alpha\gamma}\right)
^{2}}\text{ \ .} \label{EQ_crossgposi}%
\end{equation}
in the isotropic case, $\Omega_{d}$ being the surface of the $d$-dimensional
unit sphere. The parameter $g$ controls the relative importance of quartic and
quadratic terms. For large $g$ the usual BCS results are recovered while the
result Eq.(\ref{Heattri}) is obtained in the opposite limit $g\rightarrow0.$
Both quartic and quadratic terms are equally important when $g^{2}\sim
\alpha\gamma$ yielding to the same criterion $\left\vert T-T_{c}\right\vert
\sim g^{2}/a\gamma$ for the borderline between regions I and III than between
regions II and III obtained above.

\section{Small stiffness for both isotropic/anisotropic cases}

We now consider the model (\ref{EQ_our_model}) in the regime $\left\vert
T-T_{c}\right\vert \gg g^{2}/a\gamma$ where the most salient features are expected.

\textit{Approaching the tricritical point from the line defined by }$h/2\pi
T=0.3$\textit{, see Fig.\ref{FIG_phase_diagram}}, one has $g=0$ and the
spectrum of the fluctuations
\begin{equation}
\varepsilon_{\mathbf{k}}^{\ast}=a\left(  T-T^{\ast}\right)  +\underset{i,j=1}{%
{\displaystyle\sum^{d}}
}\gamma_{ij}k_{i}^{2}k_{j}^{2} \label{EQ_spectrum_g_0}%
\end{equation}
becomes purely quartic inducing the following anomalous temperature
dependences of specific heat and paraconductivity:
\begin{equation}
C^{\ast}=k_{B}T^{\ast2}B_{d}\left(  \frac{a}{\gamma}\right)  ^{d/4}f_{1}%
(\eta)(T-T^{\ast})^{(d-8)/4} \label{Heattri}%
\end{equation}
with $B_{d}=3\sqrt{2}/16,1/16,1/16\sqrt{2}\pi$ for $d=1,2,3$ respectively,
and
\begin{equation}
\sigma^{\ast}=\frac{\pi}{4}\frac{e^{2}}{\hslash}k_{B}T^{\ast}C_{d}\left(
\frac{a}{\gamma}\right)  ^{(d-2)/4}f_{2}(\eta)\left(  T-T^{\ast}\right)
^{\left(  d-6\right)  /4} \label{Condtri}%
\end{equation}
with $C_{d}=3\sqrt{2}/32,1/8\pi,5/96\sqrt{2}\pi$ for $d=1,2,3$ respectively.
The functions $f_{1}(\eta)$ and $f_{2}(\eta)$ are weakly dependent on the
anisotropy factor and are equal to unity in the isotropic case $\eta=1.$
Interestingly, the $d$-dependent exponents differ from those which are typical
of the uniform superconductivity (region I on Fig.\ref{FIG_phase_diagram}) and
of the FFLO transition at larger $g$ (region II).

Consequently, the regime $\left\vert T-T_{c}\right\vert \gg g^{2}/a\gamma$
(region III) and the related crossovers with regions I and II should reveal
the existence of the tricritical point and FFLO phase. While they were derived
along the line $g=0,$ the very unusual power laws $C\sim(T-T^{\ast})^{\left(
d-8\right)  /4}$ and $\sigma\sim(T-T^{\ast})^{\left(  d-6\right)  /4}$ must
emerge independently on the way the TCP is approached.

\textit{Following the normal/FFLO transition line }$T_{c}(h)$ \textit{to reach
the TCP}, one has $g\sim\tau^{1/2\text{ }}$where $\tau=a(T^{\ast}-T)$. We
shall demonstrate that the expressions (\ref{Heatiso},\ref{Condiso}%
,\ref{HeatAniso},\ref{CondAniso}) applying in region II yield naturally the
power laws (\ref{Heattri},\ref{Condtri}) accounting for the region III when
$T\rightarrow T^{\ast}$. First from Eqs.(\ref{Heatiso},\ref{HeatAniso}), the
specific heat is proportional to the product $g^{(d-2)/2}\tau^{-3/2}$ in the
isotropic model and to $g^{-d/2}$ $\tau^{\left(  d-4\right)  /2\text{ }}$in
presence of sensible anisotropy, since the eigenvalues $\lambda_{i}$ of the
effective mass tensor scale as $g$ and $\det\lambda\sim g^{d}.$ Taking in mind
that $g\sim\tau^{1/2}$ along the transition line, one obtains the more
divergent $(T^{\ast}-T)^{(d-8)/4}$ behavior of the specific heat already
predicted as Eq.(\ref{Heattri}) in region III for both isotropic and
anisotropic models. We have used the fact that the FFLO and the second order
uniform transition lines intersect themselves at the tricritical point.
Furthermore similar scaling on Eqs.(\ref{Condiso},\ref{CondAniso}) reveals
that the conductivity in region II is proportional to $g^{d/2}\tau^{-3/2}$ for
the isotropic model and to $g^{(2-d)/2}\tau^{\left(  d-4\right)  /2}$ for
anisotropic models. In both situations, the substitution $g\sim\tau^{1/2}$
leads to the same power ($T^{\ast}-T)^{\left(  d-6\right)  /4}$, predicted
above as Eq.(\ref{Condtri}), for the paraconductivity. In summary and more
generally, quartic terms in $\varepsilon_{\mathbf{k}}$ dominate the quadratic
ones when $\left\vert T-T_{c}\right\vert \gg g^{2}/a\gamma,$ and $\left\vert
T-T_{c}\right\vert \ll T_{c}$, see region III in Fig.1.

\section{Orbital effect}

Up to now the orbital effect was neglected to emphasize the influence of the
Zeeman pair-breaking effect on the superconducting fluctuations. This is exact
in ultracold fermionic atoms since there is no orbital effect at all for
neutral objects. This is also a good approximation in superconductors with
large Maki parameter which are the ones where the FFLO state is expected. Such
a situation is generally encountered for i) layered superconductors under in
plane magnetic field when the weakness of the interplane coupling quenches the
orbital motion of the electrons (or for thin films of 3D superconductors), and
ii) magnetic superconductors where the field originates from internal ordered
magnetic moments.

Beside these favorable situations, one should evaluate the effect of orbital
pair-breaking on the results obtained previously, especially for three
dimensional superconductors under external magnetic field. Under strong
magnetic field, the electronic motion is described by the lowest Landau level
wherein the kinetic energy is totally quenched except along the direction of
the field. This provides a realization of an effective one-dimensional system.

Coming from the BCS side with $g>0$ we evaluate how the fluctuation specific
heat and paraconductivity scale as one approaches the tricritical point along
the critical line $T_{c}(h)$, namely in the limit $g\rightarrow0^{+}.$ For
strong enough fields or sufficiently close to the transition line, the main
contribution to the specific heat is given by the lowest Landau level as
\begin{equation}
C=a^{2}k_{B}T_{c}^{2}\left(  \dfrac{2eB}{\hslash}\right)  \int_{-\infty
}^{\infty}\frac{dk_{z}\text{ }}{\left(  \tau+\frac{1}{2}\hslash\omega
_{c}+gk_{z}^{2}\right)  ^{2}}\text{\ }%
\end{equation}
transforming the three-dimensional fluctuations into one dimensional
fluctuations characterized by $C=g^{-1/2}\tau_{c}^{-3/2}$, $\hslash\omega
_{c}=4geB/\hslash$ being the Landau level spacing and $\tau_{c}=\tau
+\hslash\omega_{c}/2=0$ yielding the transition line under magnetic field. In
a similar manner one may demonstrate that the magnetoconductivity along $z
$-axis is given by $\sigma_{zz}\sim g^{1/2}\tau_{c}^{-3/2}$. In sum one
encounters a divergency of the specific heat while the magnetoconductivity
$\sigma_{zz}$ is not singular in the limit $g\rightarrow0^{+}$ performed along
the BCS transition line.

Approaching the critical point along the line $g=0$ provides an interesting
field-induced crossover. In the isotropic model the specific heat is then
given by
\begin{align}
C  &  =a^{2}k_{B}T_{c}^{2}\dfrac{2eB}{\hslash}\int_{-\infty}^{\infty}%
\frac{dk_{z}\text{ }}{\left[  \tau+\gamma\left(  \frac{2eB}{\hslash}+k_{z}%
^{2}\right)  ^{2}\right]  ^{2}}\text{\ ,}\\
&  =\dfrac{2eB}{\hslash}\frac{a^{2}k_{B}T_{c}^{2}}{\gamma^{1/4}\tau^{\ast7/4}%
}\int_{-\infty}^{\infty}\frac{dx_{{}}\text{ }}{\left(  1+x^{4}+\zeta
x^{2}\right)  ^{2}}%
\end{align}
where $\tau^{\ast}=\tau+\gamma(2eB/\hslash)^{2}.$ The dimensionless parameter
$\zeta=4(eB/\hslash)\sqrt{\gamma/\tau^{\ast}}$ controls the relative
importance of quartic and quadratic terms yielding a crossover field $B^{\ast
}=$ $(\hslash/e)\sqrt{\tau^{\ast}/\gamma}$. A similar crossover between
quartic and quadratic momentum dependence was already encountered at the
borderlines of the region III, see Eq.(\ref{EQ_crossgposi}). Here the
crossover is tuned by the magnetic field along the line $g=0$. Similarly the
conductivity
\begin{equation}
\sigma_{zz}\text{\ }\sim\frac{e^{2}}{\hslash}\gamma^{2}\frac{\gamma^{1/4}%
}{\tau^{\ast5/4}}\int_{-\infty}^{\infty}\frac{x^{6}dx\text{ }}{\left(
1+x^{4}+\zeta x^{2}\right)  ^{3}}%
\end{equation}
behaves as $\tau^{\ast-5/4}$ for fields below $B^{\ast}=$ $\hslash(\tau^{\ast
}/\gamma)^{1/2}/e$ and $\tau^{\ast-3/2\text{ }}$ for larger fields.

\section{Ginzburg-Levanyuk criterion}

Now let us address the issue of the validity of the Gaussian approximation. It
is well known that Gaussian approximation breaks down in the critical regime
wherein the interactions between different fluctuation modes of the order
parameter become sizeable. In the isotropic model, Brazovski demonstrated that
due to the critical fluctuations the transition to the non-uniform state
becomes of the first order \cite{brazovski}. Recently the same conclusion has
been obtained in the anisotropic model with isolated minima of $\varepsilon
_{\mathbf{k}}$ \cite{dalidovich_yang.2004}. However performing a simple
evaluation of the $\left\vert \Psi\right\vert ^{4}$ and $\left\vert
\Psi\right\vert ^{6}$ interaction terms contained in the nonlinear MGL
functional
\cite{buzdin_kachkachi(1997),yang_agterberg.2001,dalidovich_yang.2004}, it can
be proved that the relative width of the critical region (in temperature or
magnetic field) is given by the usual Ginzburg-Levanyuk parameter
$(T_{c}/E_{F})^{4}$, $E_{F}$ being the Fermi energy \cite{konschelle.press}.
More precisely and using our notations, the width $\tau_{G}$ of the FFLO
critical regime is given by
\begin{equation}
(T-T_{c})/T_{c}=\tau_{G}\sim\left\vert \frac{g_{0}}{g}\right\vert \left(
\frac{T_{c}}{E_{F}}\right)  ^{4}\text{ }%
\end{equation}
when the fluctuations propagate quadratically, namely when $\left\vert
g/g_{0}\right\vert >(T_{c}/E_{F})^{4/3}$ (regions I and II of
Fig.\ref{FIG_phase_diagram}). Here $g_{0}$ is the value of the coefficient $g$
far from the tricritical point. Conversely when $\left\vert g/g_{0}\right\vert
<(T_{c}/E_{F})^{4/3}$ (region III of Fig.\ref{FIG_phase_diagram}) the critical
region is somewhat broadened as
\begin{equation}
(T-T_{c})/T_{c}=\tau_{G}\sim\left(  \frac{T_{c}}{E_{F}}\right)  ^{8/3}\text{ }%
\end{equation}
owing to the presence of quartic soft modes (region III of
Fig.\ref{FIG_phase_diagram}). However even the largest value of the critical
region at $g=0$ is still very small, $(T_{c}/E_{F})^{8/3}\sim10^{-8}-10^{-10}%
$. Therefore the Gaussian approximation provides an excellent description of
the fluctuation phenomena which may be observed at the superconducting FFLO
transition. On the other hand the analysis
\cite{brazovski,dalidovich_yang.2004} may be relevant to the case of fermionic
cold atoms where the ratio $T_{c}/E_{F}$ is not so small.

\section{Conclusion}

We have demonstrated that the transition towards the FFLO state reveals very
unusual and rich fluctuation regimes which may serve as a smoking gun of the
FFLO state. Atomic Fermi gases provide a unique realization of an isotropic
system while any real superconductor always presents some degree of anisotropy
owing either to crystalline effects and/or $d$-wave pairing. The fluctuation
contribution vary non-monotonously near the tricritical point when we go from
the uniform to FFLO state. In the vicinity of the tricritical point (region
III) the very special exponents are expected for specific heat and
paraconductivity divergencies owing to the presence of soft modes with unusual
quartic dispersion. The presence of the tricritical point is signaled by very
pronounced crossovers regimes which might be observed in experiments, by
varying temperature or magnetic field. Finally, the fluctuation behavior of
the magnetic susceptibility is also very peculiar at the FFLO transition and
warrants detailed analysis \cite{konschelle.press}.

\begin{acknowledgments}
We are grateful to M. H\textsc{ouzet}, J.-N. F\textsc{uchs} and K.
Y\textsc{ang} for useful comments. This work was supported by ANR Extreme
Conditions Correlated Electrons (ANR-06-BLAN-0220).
\end{acknowledgments}

\bigskip

\end{document}